\documentclass[10pt,leqno]{amsart}
\usepackage{graphicx}
\usepackage{indentfirst,csquotes}

\usepackage{amssymb,amsthm,amsmath}
\usepackage{xcolor,paralist,hyperref,titlesec,fancyhdr,etoolbox}

\titleformat{\section}[display]{\normalfont\huge\bfseries\centering}{\centering\chaptertitlename\thechapter}{10pt}{\Large}
\titlespacing*{\section}{0pt}{0ex}{0ex}

\hypersetup{ colorlinks=true, linkcolor=black, filecolor=black, urlcolor=black }

\usepackage{lipsum}

\begin{document}
\title{Como o Uso de Máscaras Ajudou no Combate a COVID Durante a Pandemia nos Anos de 2020-2022. Análise Comparativa da Estrutura de Filtragem das Máscaras} 
\author{Maurício S. Almeida}
\author{Rodrigo Q. Almeida}
\author{Francisco Rodrigo de L. Caldas}
\author{José A. Eleutério}
\author{Letícia V. Silva}

\date{\today}
\address{Departamento de Física, Instituto Federal do Ceará - campus Juazeiro do Norte}
\email{mauricio.almeida@ifce.edu.br}
\maketitle

\let\thefootnote\relax

\begin{abstract}
O uso de máscaras faciais desempenhou um papel crucial no enfrentamento da pandemia de COVID-19, especialmente quando as vacinas ainda não estavam disponíveis. Este estudo explora a capacidade desses equipamentos de conter vírus e bactérias de dimensões nanométricas, que são menores do que os poros das próprias máscaras. Três tipos de máscaras foram selecionados para análise: máscaras de pano, cirúrgicas e PFF2. Com o objetivo de comparar a estrutura de cada uma delas, conduzimos uma análise utilizando microscopia eletrônica de varredura, que nos permitiu examinar em detalhes as diferentes camadas de filtragem presentes em cada tipo de máscara.
\end{abstract}

\bigskip

\noindent

\textbf{INTRODUÇÃO}

A pandemia causada pelo novo coronavírus (Covid-19) ceifou a vida de milhões de pessoas em todo o mundo \cite{afp}, sendo mais de 700 mil vidas só no Brasil \cite{gov}. Consequentemente, encolheu a economia global \cite{FMI} além de impactar negativamente na educação, fechando escolas e universidades.

Antes de existir uma vacina eficaz para o combate do vírus, várias medidas protetivas foram adotadas pelos países, entre elas o uso de máscaras e lockdown.

Mesmo depois do período crítico foi possível observar o surgimento de novas ondas de contaminação. Com a retomada gradual das atividades de estudo e trabalho em sua forma presencial, a população presenciou ainda o surgimento de novas variantes, como a ômicron, que possuía um potencial de contaminação muito maior do que as variantes anteriores \cite{he2021sars}..

Nesse contexto, apresentamos o objetivo geral desse trabalho, que é mostrar a importância do uso de máscaras no combate a doenças durante períodos de epidemias e pandemias, bem como explicar como elas são capazes de conter vírus e pequenas partículas, que possuem um diâmetro médio consideravelmente menor do que os poros dos seus filtros.

Este trabalho é organizado na seguinte sequência: no capítulo 2, apresentamos uma breve revisão a respeito do coronavírus; no capítulo seguinte fazemos uma abordagem sobre a Microscopia Eletrônica e o procedimento experimental utilizado para analisar amostras de máscaras em um Microscópio Eletrônico de Varredura MEV; no capítulo 4 apresentamos os resultados obtidos para cada amostra analisada e em seguida, no capítulo 5, fazemos nossas considerações finais.

\bigskip
\textbf{CORONAVÍRUS}

\begin{figure}[!htb]
\centering
\includegraphics[scale=3]{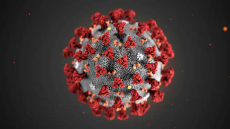}
\caption {Imagem do novo coronavírus COVID-19 em 
microscopia, fornecido pelo CDC da China.}
\label{coronavirus}
\end{figure}

Coronavirus (Fig. \ref{coronavirus}) são vírus  de RNA, tamanho de 60 nm a 140 nm, com ampla distribuição entre humanos, outros mamíferos e aves. Na microscopia eletrônica, estes vírus são vistos  com espículas que se exteriorizam de 
sua superfície, parecendo uma coroa, daí o nome corona , coroa em latim.\cite{Singhal}

Nas últimas duas décadas houve o cruzamento do vírus corona animal para humanos, isso resultou em doenças graves. A primeira vez foi em 2002, quando um novo betacoronavírus,  com origem em morcegos, atravessou para os seres humanos através do hospedeiro intermediário de gatos na província de Guangdong, China. 

Quase uma década depois, em 2012, o Coronavírus da síndrome respiratória oriental (MERS-CoV), também de origem morcego, emergiu na Arábia Saudita com camelos dromedários como hospedeiro intermediário.  

Em 2020, a Organização Mundial da Saúde nomeou o novo coronavírus como SARS-CoV-2, o causador da COVID-19, que foi identificado em pacientes com pneumonia na cidade de Wuhan, província de Hubei, China, em Dezembro de 2019. \cite{SBP}.

A COVID-19 pode se espalhar pela boca ou nariz de uma pessoa infectada em pequenas partículas líquidas quando ela tosse, espirra, fala, canta ou respira.
As evidências disponíveis atualmente sugerem que o vírus se espalha principalmente entre pessoas que estão em contato próximo umas com as outras, normalmente dentro de 1 metro (curto alcance). Uma pessoa pode ser infectada quando aerossóis ou gotículas contendo o vírus são inalados ou entram em contato direto com os olhos, nariz ou boca. O vírus também pode se espalhar em ambientes internos mal ventilados ou com aglomerações, isso ocorre porque os aerossóis permanecem suspensos no ar ou viajam a mais de 1 metro. As pessoas também podem ser infectadas ao tocar em superfícies que foram contaminadas pelo vírus e, em seguida, tocarem em seus olhos, nariz ou boca, sem limparem as mãos.
\cite{OPAS}

\bigskip
\textbf{MICROSCOPIA ELETRÔNICA}

A microscopia eletrônica de varredura (MEV) é uma técnica avançada de imagem que utiliza feixes de elétrons para visualizar a superfície de amostras com alta resolução e detalhamento. Ao contrário da microscopia óptica convencional, que utiliza luz visível para formar imagens, a MEV utiliza elétrons, que têm comprimentos de onda muito menores, permitindo uma resolução maior.

\begin{figure}[!htb]
\centering
\includegraphics[scale=0.7]{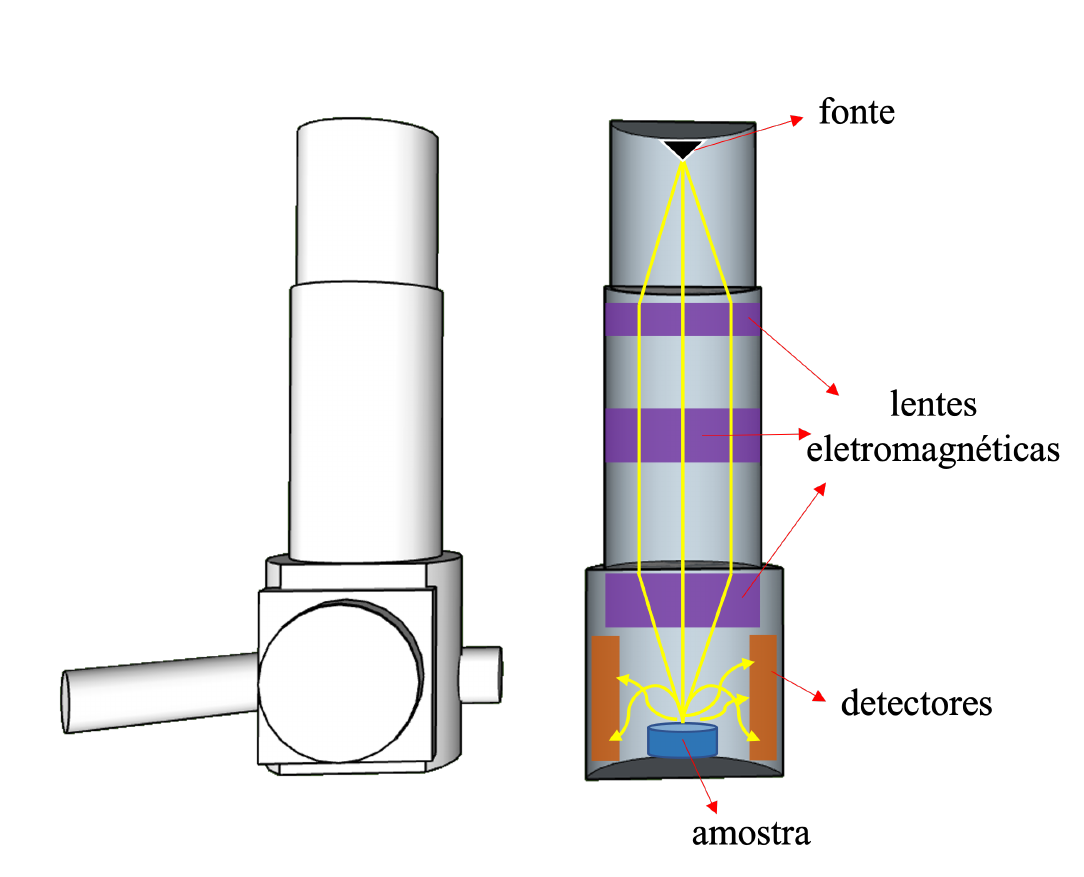}
\caption{Esquema do Microscópio Eletrônico de Varredura - MEV. Na parte superior se encontra a fonte, que consiste de um filamento aquecido responsável pela emissão de elétrons. Esses elétrons emitidos são acelerados por um conjunto de eletrodos e lentes eletromagnéticas, formando uma coluna de elétrons acelerados em direção à amostra.}
\label{mev}
\end{figure}

Dessa forma, o microscópio eletrônico de varredura utiliza um canhão de elétrons como fonte. Ela gera um feixe concentrado de elétrons de alta energia que é focalizado na superfície da amostra por meio de um conjunto de lentes eletromagnéticas (Fig. \ref{mev}) permitindo que ele atinja uma área muito pequena com alta precisão.

Quando o feixe de elétrons atinge a amostra, ele interage com a superfície e provoca a emissão de elétrons secundários e retroespalhados que são coletados por detectores específicos. A intensidade desses elétrons é mapeada em função da posição, formando uma imagem da superfície da amostra. 

Antes de ser visualizada, a amostra deve ser preparada adequadamente. Isso pode envolver a fixação, desidratação, revestimento com metais condutores (como ouro ou platina) e, em alguns casos, até mesmo o congelamento para preservar a estrutura da amostra.

Como o material utilizado na confecção das máscaras faciais não é naturalmente condutor de eletricidade, foi necessário realizar a metalização da amostra.

Esse procedimento consistiu na aplicação de uma fina camada de ouro sobre a superfície da amostra. Dessa forma, quando o feixe de elétrons atingisse a amostra, haveria a emissão de elétrons, fundamentais para a formação da imagem.

\bigskip
\textbf{PROCEDIMENTO EXPERIMENTAL}

\begin{figure}[!htb]
\centering
\includegraphics[scale=0.4]{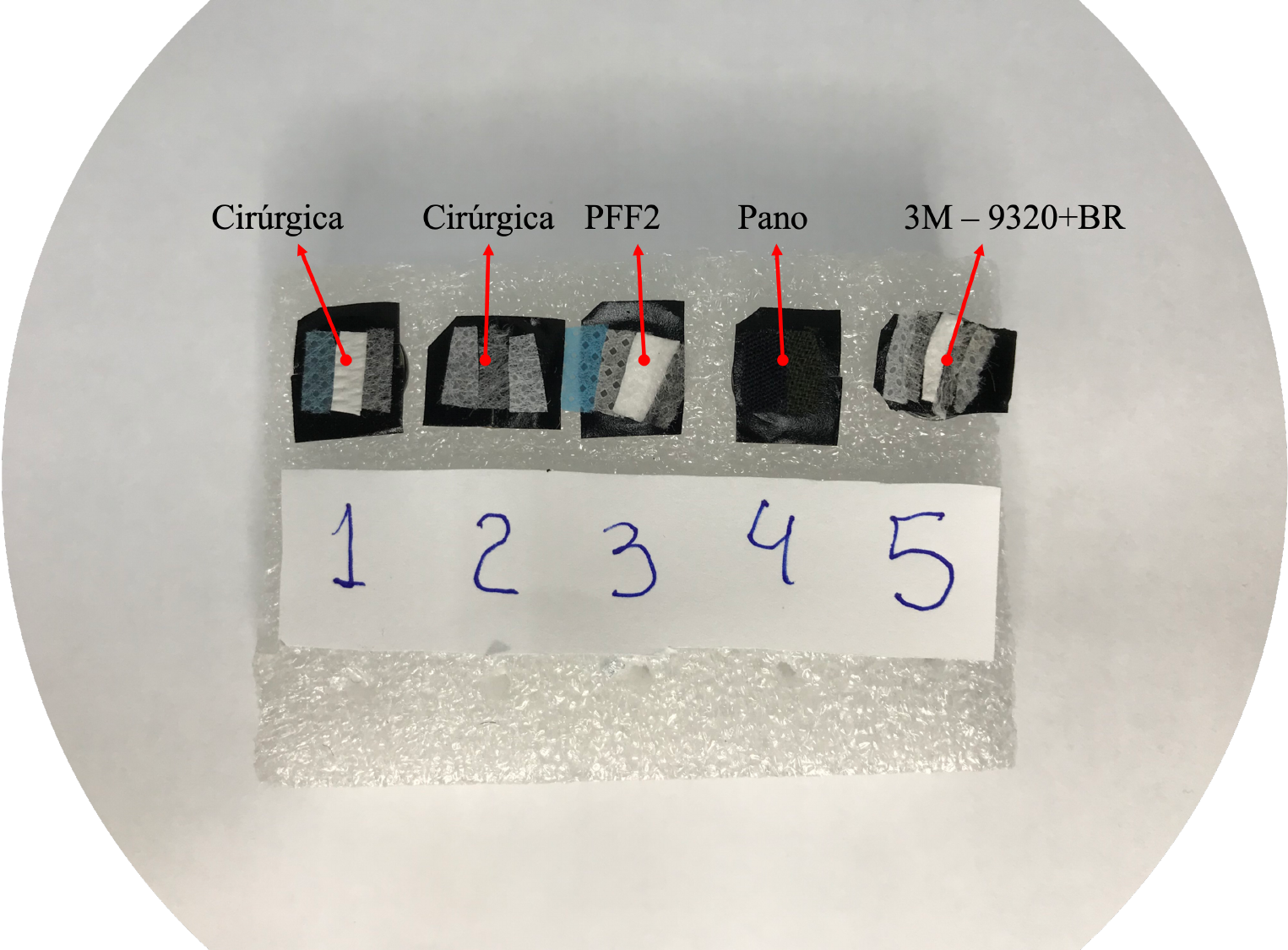}
\caption{Foram utilizadas cinco máscaras para o experimento: duas cirúrgicas, uma do tipo PFF2, uma de pano e outra também PFF2, da marca 3M, modelo 3M-9320+BR. As amostras foram cortadas, separadas suas camadas e posicionadas lado a lado nos $stubs$ para melhor análise dos resultados.}
\label{amostras}
\end{figure}

Foram analisadas três tipos de máscaras: pano, cirúrgica e PFF2, conforme representado na Fig. \ref{amostras}. Elas foram cortadas em pequenos pedaços e fixadas em $stubs$ cobertos por carbono, ocupando uma área total de aproximadamente $1,/cm^{2}$. Isso é necessário porque amostras inorgânicas e biológicas/poliméricas devem ser metalizadas para a análise MEV.

\begin{figure}[!htb]
\centering
\includegraphics[scale=0.4]{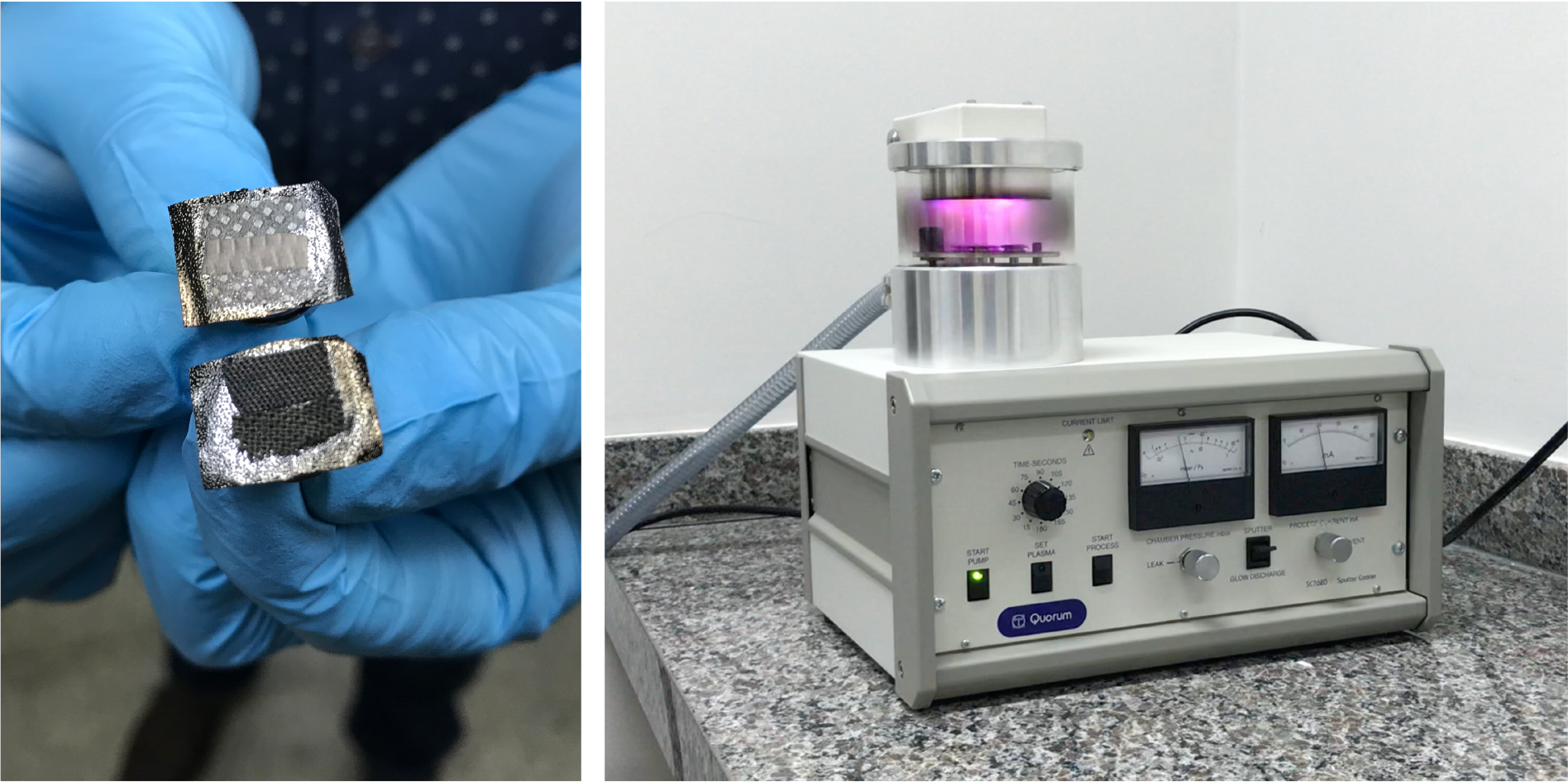}
\caption{À esquerda temos uma foto da amostra depois do processo de metalização. Sobre ela foi depositada uma fina camada de ouro. Isso foi obtido por meio da câmara de metalização, foto à direita. Nela, é possível ver a formação do plasma (nuvem roxa) durante o processo de deposição do metal.}
\label{magnetizar}
\end{figure}

Em seguida, as amostras foram levadas para uma câmara de metalização onde foi utilizado o ouro como metal de revestimento. Essa câmara foi aquecida ocasionando a vaporização do metal, conforme pode ser observado na Fig. \ref{magnetizar}, provocando a condensação do ouro sobre a superfície da amostra, criando uma fina e uniforme camada.

Após a deposição do metal, também observado na Fig. \ref{magnetizar}, a amostra foi cuidadosamente removida e inserida no microscópio eletrônico de varredura, para a obtenção das imagens.

\begin{figure}[!htb]
\centering
\includegraphics[scale=0.4]{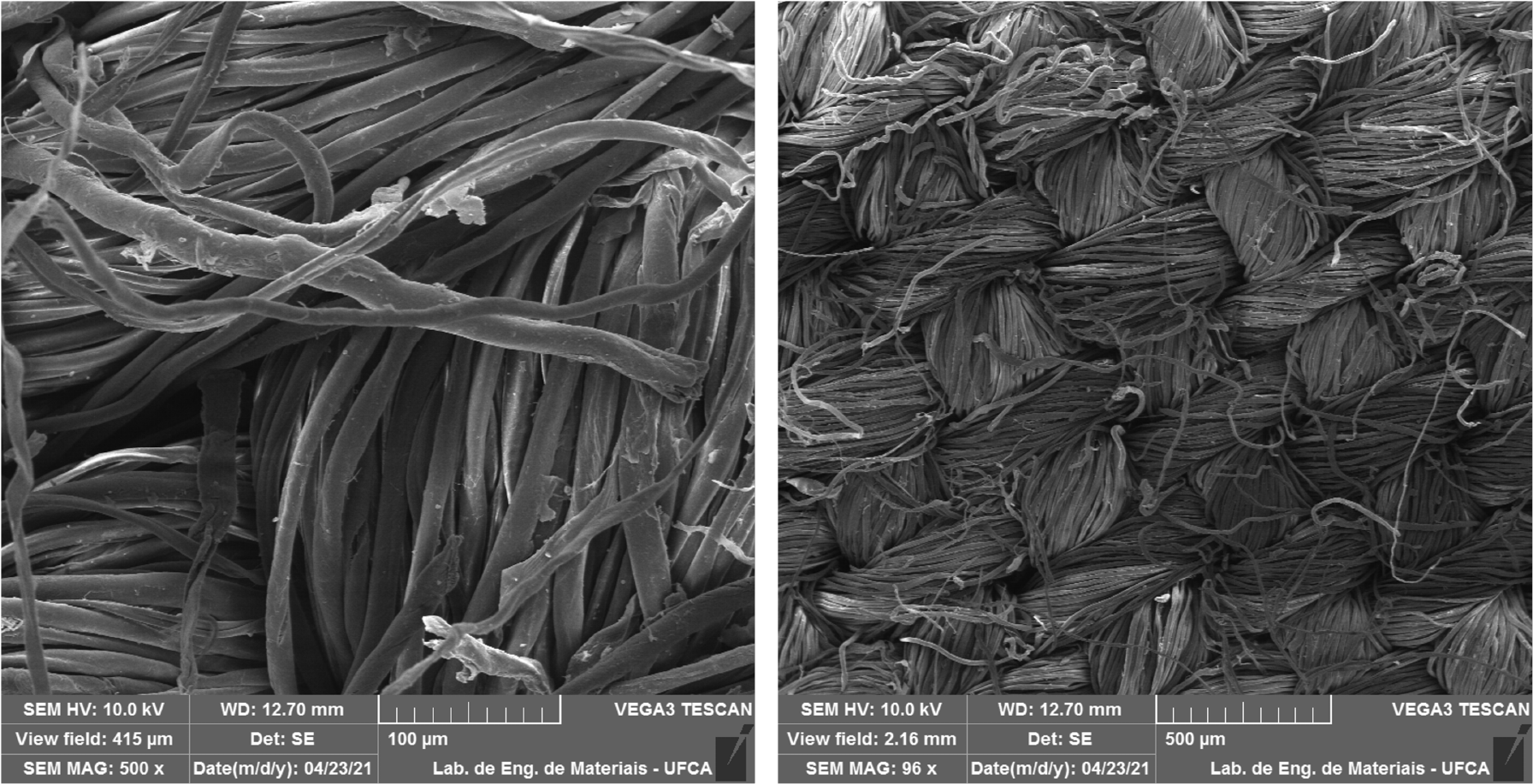}
\caption{Máscara de pano. A figura à esquerda apresenta um campo de visualização de 415\, $\mu$m, onde é possível ver os detalhes dos poros existenes entre os fios de algodão. Já na figura à direita, com um campo de visão maior de 2,16mm, podemos ver o padrão de organização dos fios do tecido.}
\label{pano}
\end{figure}

\bigskip
\textbf{DISCUSSÃO E RESULTADOS}
A Fig. \ref{pano} apresenta a amostra da máscara de pano. Nela é possível observar uma orientação bem definida nos fios de algodão, formando uma malha estruturada.

Para verificar o diâmetro médio dos poros presentes entre os fios, utilizou-se o software de análises de imagme ImajeJ. O procedimento consistiu em delimitar uma área de 10\,$mm^{2}$ e contar a quantidade de poros que ocupavam aquela região. Em seguida, foi contabilizado o diâmetro médio de cada um desses poros. 

\begin{figure}[!htb]
\centering
\includegraphics[scale=0.50]{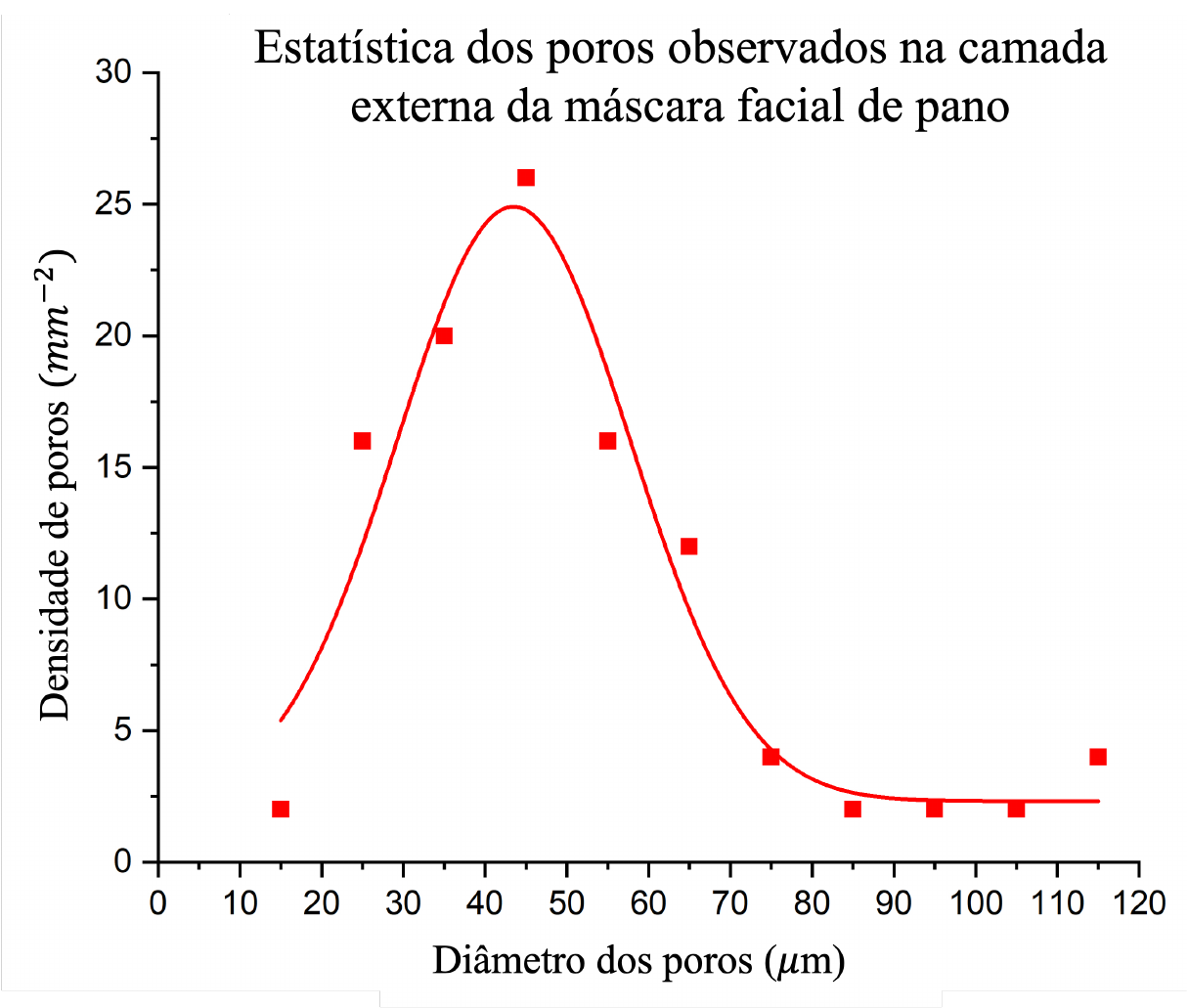}
\caption{O eixo vertical apresenta a densidade de poros, dada pela razão entre o número de poros por área de 10\,$mm^{2}$. É possível observar que a maioria dos poros possuem diâmetro compreendidos entre uma faixa de 30 a 60\,$\mu$m, tamanho consideravelmente maior dos que as dimensões do coronavírus. Os dados desse gráfico foram obtidos utilizando o software de análise de imagens ImageJ.}
\label{poro_pano}
\end{figure}

Para a máscara de pano, que tem fios de algodão com 15\,$\mu$m de espessura, observou-se poros com diâmetros, predominantemente, entre 30\,$\mu$m e 60\,$\mu$m, conforme é possível ver na Fig. \ref{poro_pano}.

Acontece que, como foi mencionado na Seção 2, o coronavírus é muito menor que todas essas dimensões. Ele tem diâmetro em uma faixa que vai de 60\,nm a 140\,nm; isso é cerca de mil vezes menor do que o diâmetro de um fio de cabelo, o que faria que, em tese, pudesse atravessar facilmente uma máscara de pano.

Entretanto, é preciso lembrar que vírus e bactérias são transmitidos pelo ar em gotículas que possuem uma ampla variedade de tamanhos, de tal forma que as maiores acabam sendo retidas pelas máscaras, mesmo que sejam de pano.

As menores dessas partículas são chamadas de aerossóis. Estes são os principais propagadores de contaminação e conseguem deslizar com facilidade por aberturas de certas fibras de tecido.

Sendo assim, como as máscaras de pano não obedecem a um padrão de confecção, elas acabam tendo um nível de eficácia variado. Entretanto, mesmo com essa limitação, podem fornecer alguma proteção a terceiros, caso estejam sendo utilizadas por pessoas infectadas, principalmente se a máscara tiver mais de uma camada de tecido.

\begin{figure}[!htb]
\centering
\includegraphics[scale=0.4]{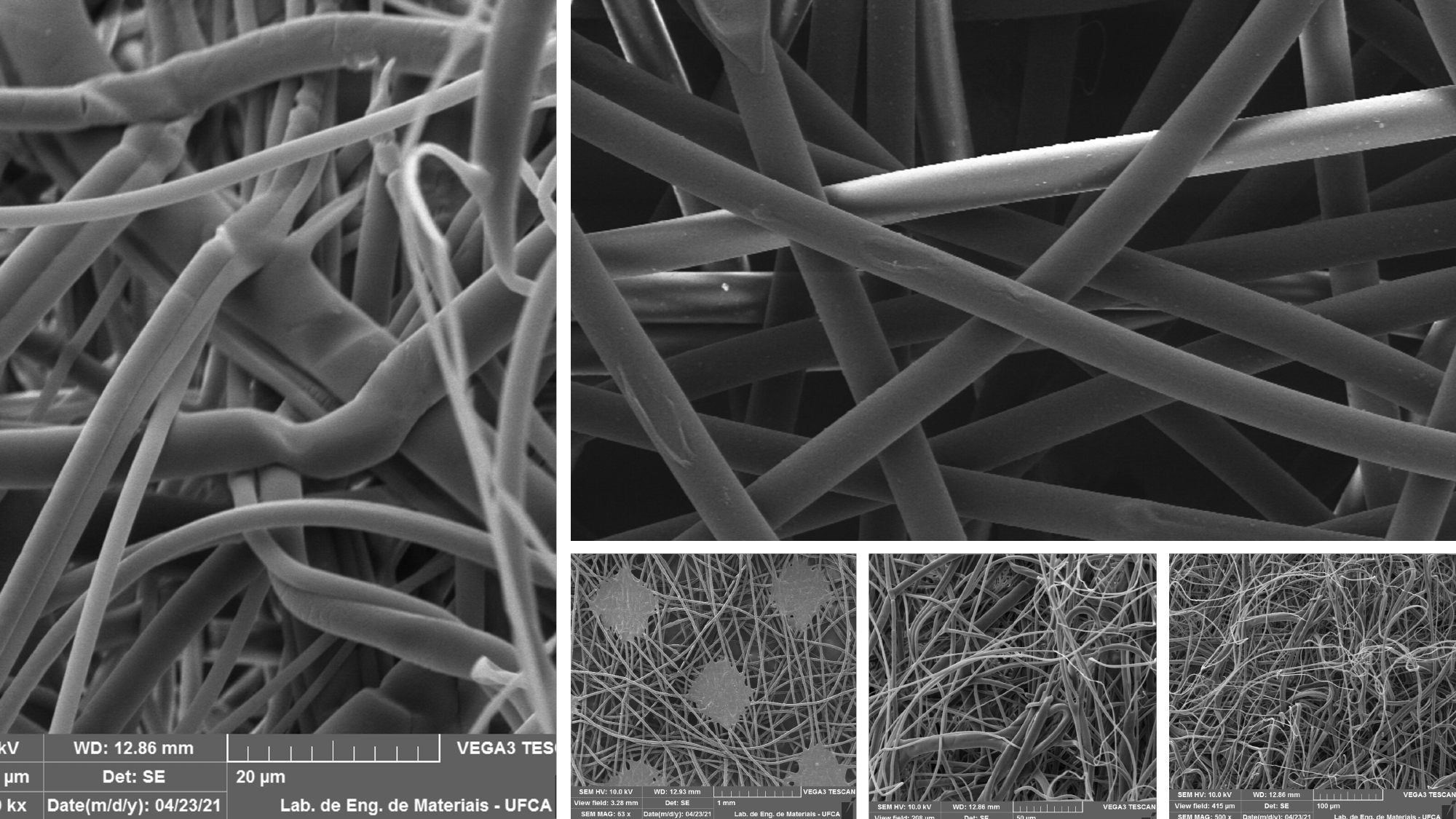}
\caption{Máscara cirúrgica. As três imagens no lado inferior direito da figura mostram as três camadas da máscara cirúrgica. É possível verificar mas marcas deixadas no processo de termofixação (esquerda) e os detalhes da camada filtrante (direita). As duas imagens maiores mostram exatamente o detalhe dessa camada filtrante. Trata-se de uma rede de filamentos sem orientação definida.}
\label{cirurgica}
\end{figure}

As máscaras cirúrgicas (Fig. \ref{cirurgica}), são compostas por três camadas, sendo a primeira de TNT, a do meio de um material filtrante e a última, que fica em contato com boca e nariz, também é de um material filtrante que tem afinidade com a água.

Na imagem é possível observar a estrutura do TNT que é produzido através da deposição e termofixação de fibras de polipropileno não orientadas, numa rede sem orientação preferencial aparente e com porosidade média menor do que a apresentada pela máscara de pano.

Também na Fig. \ref{cirurgica} é possível observar áreas que são resultado do processo de fabricação, em que as fibras são fixadas termicamente a pressão com temperaturas elevadas.

É válido lembrar que as camadas de filtração da máscara facial devem possuir um diâmetro de poro médio pequeno o suficiente para prender e filtrar as partículas; e grande o suficiente para permitir uma respiração confortável.

Entretanto, não são as menores partículas que preocupam, pois aerossóis com diâmetro menor ou igual a 0,1\,$\mu$m descrevem o chamado movimento Browniano, que é um movimento aleatório. Dessa forma, a probabilidade que em algum momento esses aerossóis menores encontrem uma fibra e fiquem presos a elas é muito grande.

O maior perigo acontece com os aerossóis de tamanho intermediário, com diâmetros entre 0,1\,$\mu$m a 1\,$\mu$m, pois estes conseguem contornar os obstáculos ao longo do escoamento, similarmente a partículas de fumaça utilizadas em túneis de vento para traçarem as linhas de corrente.

\begin{figure}[!htb]
\centering
\includegraphics[scale=0.4]{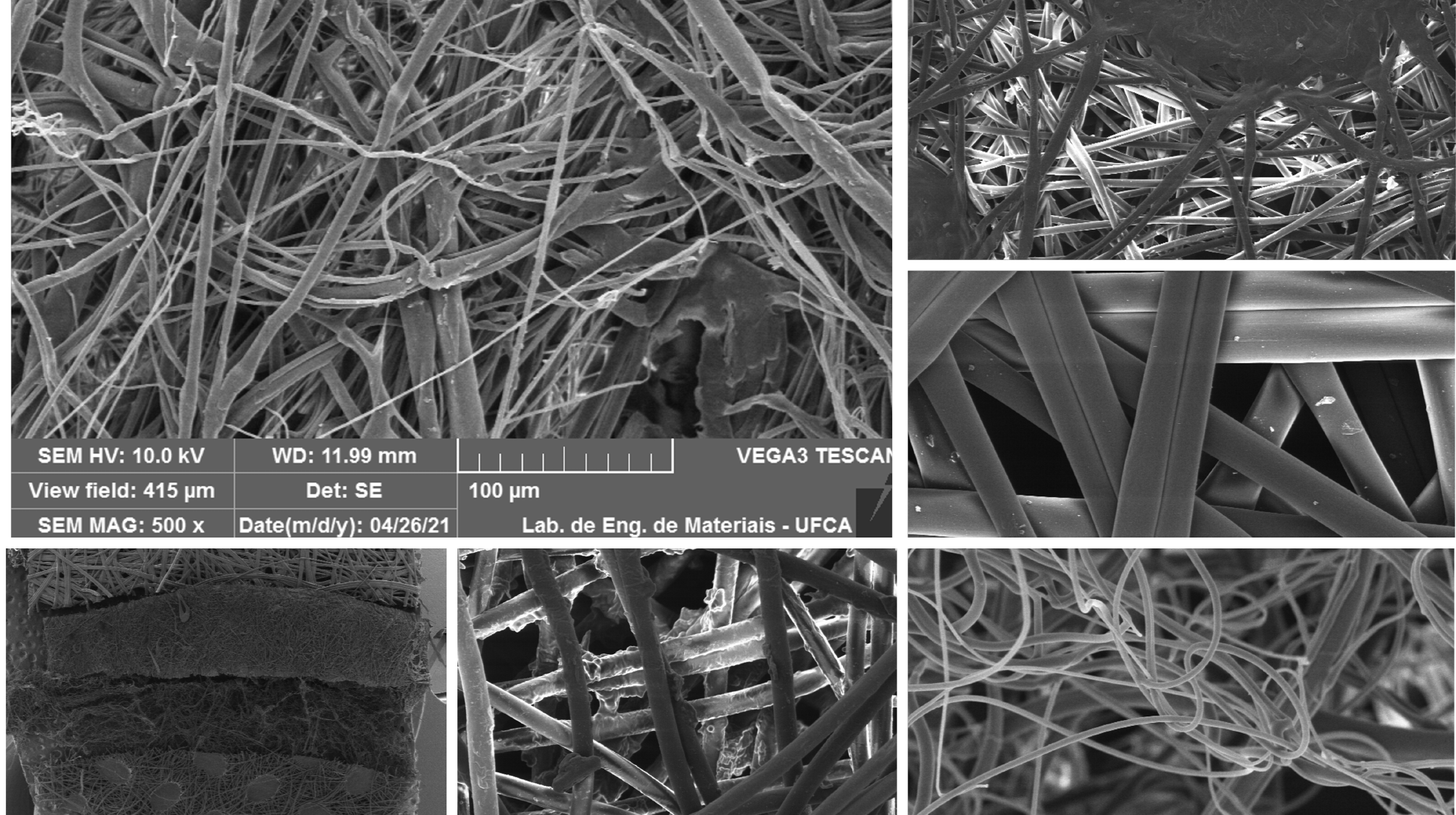}
\caption{Máscara 3M. Essa máscara conta com uma dupla camada de filtragem (uma delas apresentada na foto maior), além de uma camada eletrostática capaz de capturar partículas de tamanho intermediário, com diâmetros entre 0,1\,$\mu$m a 1\,$\mu$m, que são as mais difíceis de serem retidas.}
\label{3m}
\end{figure}

É aqui que entra a importância das máscaras do tipo PFF2 (Fig. \ref{3m}), também chadada de N95. Isso porque, para serem capazes de filtrar também esses essas partículas intermediárias, as máscaras do tipo PFF2 possuem uma camada eletrostática, incorporada entre as camadas de filtragem da máscara padrão, fazendo com que partículas sejam atraídas pela força de Coulomb, de longo alcance, em direção a camada eletrostática. 

Uma vez capturadas, as partículas são mantidas no lugar por meio das forças de Van der Walls. O nome N95 quer dizer exatamente isso, que a máscara consegue filtrar pelo menos 95$\%$ das partículas de tamanho intermediário, as mais difíceis de segurar.

A máscara 3M-9320-BR também contava com 5 camadas de proteção (Fig. \ref{3m}), contendo uma de TNT, duas filtrantes, uma eletrostática e a última, que fica em contato com a boca, do tipo hidrófila. 

Além disso, esse tipo de máscara tem melhor capacidade de vedação, sendo capaz de proteger não apenas o usuário de inalar partículas contaminadas, como de proliferar contaminação (caso esteja infectado) para terceiros que estejam próximos.

\bigskip
\textbf{CONCLUSÃO}

O uso de máscaras foi uma das medidas-chave adotadas para combater a propagação do COVID durante o período da pandemia. 

Como os vírus são transportados em pequenas gotículas liberadas quando pessoas infectadas tossem, espirram, falam ou respiram, as máscaras faciais, nesse contexto, funcionam como uma barreira física que capturam essas gotículas, evitando que elas se dispersem no ar e contaminem outras pessoas.

Mesmo as máscaras de pano, que apresentaram diâmetros de poros entre 30\,$\mu$m e 60\,$\mu$m, também podem ser usadas como barreiras física. Entretanto, por não terem o mesmo padrão de confecção, apresentam nível de eficácia variável.

Em contrapartida, as máscaras cirúrgicas são formadas por uma tripla camada de proteção, onde a mais externa é repelente a fluidos; a intermediária tem uma característica filtrante, capaz de reter gotículas portadoras dos vírus; e a camada que fica em contato com a boca possui característica hidrofílica, que permite a absorção de secreções salivares.

Além disso, a disposição aleatória dos fios da camada intermediária serve para capturar vírus e pequenas partículas que executam movimento Browniano, uma vez que essas partículas tendem a colidir e ser capturadas pelas fibras e superfícies da camada de filtro durante esse movimento caótico.

A máscara do tipo PFF2 conta ainda com uma camada extra de filtragem, eletrostática, que faz com que as partículas intermediárias, que não executam movimento Browniano e que possuem diâmetro menor do que os poros da camada de filtro, sejam capturadas por meio de força eletrostática. Além de possuir uma estrutura com poder de filtragem superior, esse tipo de máscara oferece melhor vedação.

A máscara 3M-9320-BR foi a que apresentou mais camadas de proteção, dentre as amostras analisadas, além de boa vedação. Ela contava com duas camadas internas filtrantes, uma camada eletrostática, além da parte externa recoberta por um não tecido, que serve de proteção ao meio filtrante impedindo que as fibras se soltem. 

\bigskip
\textbf*{Agradecimentos}
Os autores agradecem ao departamento de Engenharia de Materiais da Universidade Federal do Cariri - UFCA, pela realização das análises de microscopia das amostras.

\end{document}